\documentclass[aps,prd,twocolumn,showpacs,nofootinbib]{revtex4-1}

\usepackage{amssymb} \usepackage{amsmath} \usepackage{graphicx}
\usepackage{epsfig,latexsym}
\usepackage{mathtools}

\RequirePackage{xspace} \allowdisplaybreaks

\begin{document}

\def\bef{\begin{figure}}
\def\eef{\end{figure}}

\newcommand{\nl}{\nonumber\\}

\newcommand{\ans}{ansatz }
\newcommand{\be}[1]{\begin{equation}\label{#1}}
\newcommand{\beq}{\begin{equation}}
\newcommand{\ee}{\end{equation}}
\newcommand{\beqn}[1]{\begin{eqnarray}\label{#1}}
\newcommand{\eeqn}{\end{eqnarray}}
\newcommand{\bd}{\begin{displaymath}}
\newcommand{\ed}{\end{displaymath}}
\newcommand{\mat}[4]{\left(\begin{array}{cc}{#1}&{#2}\\{#3}&{#4}
\end{array}\right)}
\newcommand{\matr}[9]{\left(\begin{array}{ccc}{#1}&{#2}&{#3}\\
{#4}&{#5}&{#6}\\{#7}&{#8}&{#9}\end{array}\right)}
\newcommand{\matrr}[6]{\left(\begin{array}{cc}{#1}&{#2}\\
{#3}&{#4}\\{#5}&{#6}\end{array}\right)}
\newcommand{\cvb}[3]{#1^{#2}_{#3}}
\def\lsim{\raise0.3ex\hbox{$\;<$\kern-0.75em\raise-1.1ex
e\hbox{$\sim\;$}}}
\def\gsim{\raise0.3ex\hbox{$\;>$\kern-0.75em\raise-1.1ex
\hbox{$\sim\;$}}}
\def\abs#1{\left| #1\right|}
\def\simlt{\mathrel{\lower2.5pt\vbox{\lineskip=0pt\baselineskip=0pt
           \hbox{$<$}\hbox{$\sim$}}}}
\def\simgt{\mathrel{\lower2.5pt\vbox{\lineskip=0pt\baselineskip=0pt
           \hbox{$>$}\hbox{$\sim$}}}}
\def\unity{{\hbox{1\kern-.8mm l}}}
\newcommand{\eps}{\varepsilon}
\def\ep{\epsilon}
\def\ga{\gamma}
\def\Ga{\Gamma}
\def\om{\omega}
\def\omp{{\omega^\prime}}
\def\Om{\Omega}
\def\la{\lambda}
\def\La{\Lambda}
\def\al{\alpha}
\newcommand{\ov}{\overline}
\renewcommand{\to}{\rightarrow}
\renewcommand{\vec}[1]{\mathbf{#1}}
\newcommand{\vect}[1]{\mbox{\boldmath$#1$}}
\def\tm{{\widetilde{m}}}
\def\mcirc{{\stackrel{o}{m}}}
\newcommand{\Dm}{\Delta m}
\newcommand{\dm}{\varepsilon}
\newcommand{\tanb}{\tan\beta}
\newcommand{\nbar}{\tilde{n}}
\newcommand\PM[1]{\begin{pmatrix}#1\end{pmatrix}}
\newcommand{\up}{\uparrow}
\newcommand{\down}{\downarrow}
\def\omE{\omega_{\rm Ter}}

%

\newcommand{\Dsusy}{{susy \hspace{-9.4pt} \slash}\;}
\newcommand{\DCP}{{CP \hspace{-7.4pt} \slash}\;}
\newcommand{\mc}{\mathcal}
\newcommand{\gr}{\mathbf}
\renewcommand{\to}{\rightarrow}
\newcommand{\gtc}{\mathfrak}
\newcommand{\wh}{\widehat}
\newcommand{\br}{\langle}
\newcommand{\kt}{\rangle}


\def\lsim{\mathrel{\mathop  {\hbox{\lower0.5ex\hbox{$\sim$}
\kern-0.8em\lower-0.7ex\hbox{$<$}}}}}
\def\gsim{\mathrel{\mathop  {\hbox{\lower0.5ex\hbox{$\sim$}
\kern-0.8em\lower-0.7ex\hbox{$>$}}}}}

\def\nn{\\  \nonumber}
\def\de{\partial}
\def\brf{{\mathbf f}}
\def\bbf{\bar{\bf f}}
\def\bF{{\bf F}}
\def\bbF{\bar{\bf F}}
\def\bA{{\mathbf A}}
\def\bB{{\mathbf B}}
\def\bG{{\mathbf G}}
\def\bI{{\mathbf I}}
\def\bM{{\mathbf M}}
\def\bY{{\mathbf Y}}
\def\bX{{\mathbf X}}
\def\bS{{\mathbf S}}
\def\bb{{\mathbf b}}
\def\bh{{\mathbf h}}
\def\bg{{\mathbf g}}
\def\bla{{\mathbf \la}}
\def\bmu{\mathbf m }
\def\by{{\mathbf y}}
\def\bmu{\mbox{\boldmath $\mu$} }
\def\bsig{\mbox{\boldmath $\sigma$} }
\def\bunity{{\mathbf 1}}
\def\cA{{\cal A}}
\def\cB{{\cal B}}
\def\cC{{\cal C}}
\def\cD{{\cal D}}
\def\cF{{\cal F}}
\def\cG{{\cal G}}
\def\cH{{\cal H}}
\def\cI{{\cal I}}
\def\cL{{\cal L}}
\def\cN{{\cal N}}
\def\cM{{\cal M}}
\def\cO{{\cal O}}
\def\cR{{\cal R}}
\def\cS{{\cal S}}
\def\cT{{\cal T}}
\def\eV{{\rm eV}}

%

\title{(Anti)evaporation of Dyonic Black Holes in string-inspired  
dilaton $f(R)$-gravity 
}

\author{Andrea Addazi$^1$}\email{andrea.addazi@infn.lngs.it}
\affiliation{$^1$ Dipartimento di Fisica,
 Universit\`a di L'Aquila, 67010 Coppito AQ and
LNGS, Laboratori Nazionali del Gran Sasso, 67010 Assergi AQ, Italy}

\begin{abstract}

We discuss dyonic black hole solutions
in the case of $f(R)$-gravity 
coupled with a dilaton and 
two gauge bosons. 
The study of such a model is highly motivated from string theory.
Our Black Hole solutions are extensions 
of the one firstly studied by 
Kallosh, Linde, Ort\'in, Peet and Van Proyen
(KLOPV) in [arXiv:hep-th/9205027].
We will show that extreme solutions are unstable.
In particular, these solutions 
have
Bousso-Hawking-Nojiri-Odintsov
(anti)evaporation instabilities.

\end{abstract}

\maketitle
\section{Introduction}

As is known, 
the low energy limit of a dimensionally reduced superstring 
theory to $d=4$
is $\mathcal{N}=4$ supergravity. 
There are two versions:
$SO(4)$ and $SU(4)$. 
The first one is invariant
under a (rigid) $SU(4)\times SU(1,1)$ symmetry. 
Black hole solutions of the reduced sector $U(1)^{2}$ 
were studied  by Kallosh, Linde, Ort\'in,
Peet and Van Proeyen (KLOPV)
in 
Ref. \cite{Kallosh:1992ii}.
In particular, they
consider 
$U(1)^{2}$ charged dilaton black holes.
These solutions are
Reissner-Nordstr\"om-like black holes,
or more precisely of dyonic black holes. 
In particular, the dilaton field is the real part 
of an initial complex scalar, while the 
imaginary part is an axion pseudoscalar field. 
They assumed the axion stabilized to a constant VEV. 
The effective bosonic action 
corresponds to the Einstein-Hilbert one
coupled with a dilaton field and 
two $U(1)$ fields.
Extreme limits of dyonic solutions are shown to saturate $\mathcal{N}=4$
supersymmetry in $d=4$. 
On the other hand, the presence of non-perturbative stringy effects 
could modify the effective action in the low energy limit. 
For instance, higher derivative terms may be generated by Euclidean D-brane or worldsheet instantons. 
In particular, the Einstein-Hilbert sector coupled to the dilaton and $U(1)$-fields can be 
extended from $R$ to an analytic function $f(R)$   (See Ref.\cite{Bianchi:2009ij} for a review on this subject)
\footnote{See Refs.\cite{Addazi:2015goa,Addazi:2015yna,Addazi:2016mtn,Addazi:2016xuh}
for recent investigations of E-brane instantons in particle physics. }. 

KLOPP solutions are particularly important in string theory. 
For instance the famous 
derivation of the Hawking BH entropy
from BPS microstates shown by 
 Strominger and Vafa 
is based on five dimensional KLOPP solutions \cite{Strominger:1996sh}.
The Vafa-Strominger result has inspired the 
so called fuzzball proposal,
which has the ambition to solve the BH information 
paradox \cite{Mathur:2009hf}.

In this paper, we will study black hole solutions 
in string inspired $f(R)$-gravity, coupled with a dilaton field and two gauge bosons. 
We assume that the asymptotic space-time is 
Minkowski's one. 
Let us clarify that we will not consider a $f(R)$-{\it supergravity} coupled
to gauge bosons and dilatons. 
In fact, it was recently shown that 
the only $f(R)$-supergravity which 
is not plagued by ghosts and tachyons is 
Starobinsky's supergravity
\cite{Ketov:2013dfa,Ferrara:2013pla}. 
Nevertheless, one can consider the case in which
higher derivative terms are generated 
by exotic instantons or fluxes after a spontaneous 
supersymmetry breaking mechanism. 
In this sense, our model, which 
has a stable vacua and it is 
not plagued by ghosts and tachyons, 
is inspired by string theory. 
Clearly, to calculate instantonic corrections 
from a realistic stringy model is, at the moment,  impossible.
We believe that this highly motivates our 
effective field theory analysis, 
in which coefficients inside the $f(R)$-functional
parametrize our ignorance about the string theory vacua. 
We will show that extreme dyonic solutions 
have Bousso-Hawking-Nojiri-Odintsov (BHNO)  
  \cite{Bousso:1997wi,Nojiri:2013su,Nojiri:2014jqa} 
(anti)evaporation instabilities.
In particular, Nojiri and Odintsov have discovered
(anti)evaporation instabilities in 
Reissner-Nordstr\"om black holes 
in $f(R)$-gravity \cite{Nojiri:2014jqa}. 
 { \it At posteriori}, our result is understood 
as a generalization of Nojiri-Odintsov calculations 
in Ref. \cite{Nojiri:2014jqa}. 
On the other hand, the peculiar thermodynamical proprieties of antievaporating solutions 
were discussed in our recent paper 
 \cite{Addazi:2016prb}.

 \begin{figure}[t]
\centerline{ \includegraphics [height=5cm,width=0.9 \columnwidth]{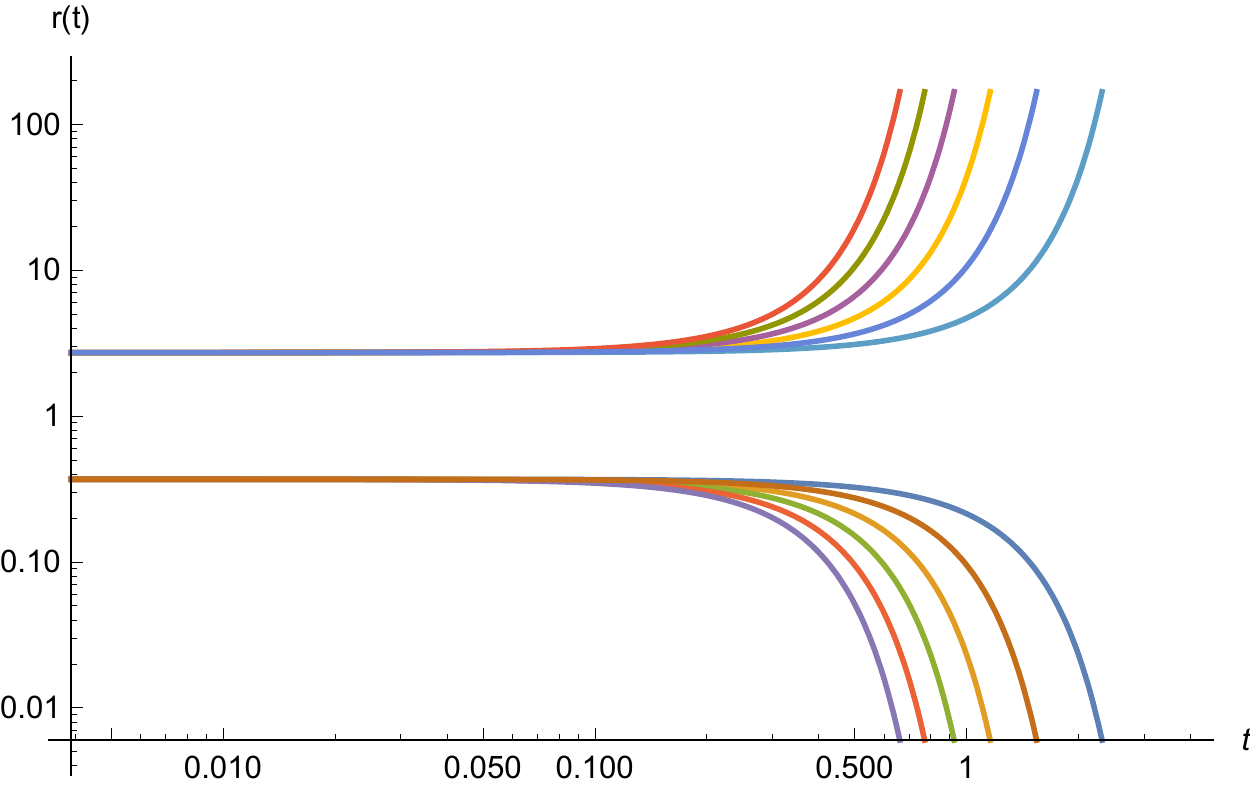}}
\vspace*{-1ex}
\caption{The evolution in time of the extremal dyonic BH horizon is displayed
imposing initial condition $\phi_{0}=\pm 1$ and $\beta=1,1.5,2,2.5,3,3.5$
(Log-Log-plot). This numerical solution is obtained from EoM 
at the second order of the perturbation theory. 
The growing and the decreasing solutions 
represent antievaporation and evaporation respectively. 
}
\label{plot}   
\end{figure}

\vspace{2cm}

\section{Dilaton-$f(R)$-gravity }

Let us consider the case of 
a $f(R)$-gravity with 
two $U(1)$-gauge bosons 
and a dilaton.
In particular, we will consider the action 
\begin{align} 
S=\int d^{4}x \sqrt{-g}[-f(R)+2\partial^{\mu}\phi \partial_{\mu}\phi \nonumber \\
+2\nabla_{\mu}\phi\nabla_{\nu}\phi-e^{-2\phi}(2F_{\mu\lambda}F_{\nu\delta}g^{\lambda\delta}-\frac{1}{2}g_{\mu\nu}F^{2})]
\end{align}
where 
$$F_{\mu\nu}=\partial_{\nu}A_{\mu}-\partial_{\mu}A_{\nu}$$
$$\tilde{B}_{\mu\nu}=\partial_{\nu}\tilde{B}_{\mu}-\partial_{\mu}\tilde{B}_{\nu}$$
and 
$A_{\mu},B_{\mu}$ are gauge bosons of $U(1)\times U(1)$, 
we conveniently use unit 
$2\kappa_{(4)}=1$, where $\kappa_{(4)}$ is the four dimensional gravitational coupling
(coming from the Kaluza-Kelin reduction of the ten-dimensional gravitational coupling). 
The action Eq.(1) comes 
 from the $SO(4)$, $d=4$, $\mathcal{N}=4$ supergravity
 and it is formulated in the Einstein-frame,
 with an opportune and understood redefinition 
 of the dilaton field.

The Equations of Motion are 
\begin{align} 
0=\nabla_{\mu}(e^{-2\phi}F^{\mu\nu}) \\
0=\nabla_{\mu}(e^{2\phi}\tilde{G}^{\mu\nu})  \\
0=\nabla^{2}\phi-\frac{1}{2}e^{-2\phi}F^{2}+\frac{1}{2}e^{2\phi}\tilde{G}^{2} \\
0=f_{R}(R)R_{\mu\nu}+\frac{1}{2}(Rf_{R}-f(R))g_{\mu\nu}\\
-\nabla_{\mu}\nabla_{\nu}f_{R}(R)+g_{\mu\nu}\Box f_{R}(R)  \nonumber \\
+2\nabla_{\mu}\phi\nabla_{\nu}\phi-e^{-2\phi}(2F_{\mu\lambda}F_{\nu\delta}g^{\lambda\delta}-\frac{1}{2}g_{\mu\nu}F^{2})\nonumber \\
-e^{2\phi}(2\tilde{G}_{\mu\lambda}\tilde{G}_{\nu\delta}g^{\lambda\delta}-\frac{1}{2}g_{\mu\nu}\tilde{G}^{2})
\end{align}

A solution of these equations is 
\begin{equation} \label{metric}
ds^{2}=e^{2U}dt^{2}-e^{-2U}dr^{2}-R^{2}d\Omega 
\end{equation}
$$e^{2\phi}=e^{2\phi_{0}}\frac{r+\Sigma}{r-\Sigma}$$
$$F=\frac{Qe^{\phi_{0}}}{(r-\Sigma)^{2}}dt \wedge dr  $$
$$\tilde{G}=\frac{Pe^{-\phi_{0}}}{(r+\Sigma)^{2}}dt \wedge dr $$
$$e^{2U}=\frac{(r-r_{+})(r-r_{-})}{R^{2}}$$
$$R^{2}=r^{2}-\Sigma^{2},\,\,\,\,\Sigma=\frac{P^{2}-Q^{2}}{2M},\,\,\,\, r_{\pm}=M\pm r_{0}$$
$$r_{0}^{2}=M^{2}+\Sigma^{2}-P^{2}-Q^{2}=M^{2}+\Sigma^{2}-e^{-2\phi_{0}}P_{m}^{2}-e^{-2\phi_{0}}Q^{2}_{el}$$

The solutions depend on independent parameters $M,Q,P,\phi_{0}$.
$M$ is the BH mass, $\phi_{0}$ is the asymptotic value of 
the dilaton field. 
$Q_{el}=e^{\phi_{0}Q}$ is the F-field electric charge, 
while 
$P_{m}=e^{\phi_{0}}P$ is the G-field magnetic charge
(electric charge of $\tilde{G}$).

These equations imply the relation 
$$Cf_{R}(R_{0})=q^{2}\equiv \sqrt{Q^{2}+P^{2}}=e^{-\phi_{0}}\sqrt{Q_{el}^{2}+P_{m}^{2}}$$
where $C$ is an integration constant. 

In the case of an extremal dyonic black hole, the
metric can be conveniently rewritten as \cite{Nojiri:2014jqa}
$$ds^{2}=\frac{M^{2}}{{\rm cosh}^{2}x}(d\tau^{2}-dx^{2})+M^{2}d\Omega^{2}$$

This suggests the ansatz 
$$ds^{2}=M^{2}e^{2\rho(x,\tau)}(d\tau^{2}-dx^{2})+M^{2}e^{-2\varphi(x,\tau)}(d\tau^{2}-dx^{2})d\Omega^{2}$$
and the gravitational EoM can be rewritten as 
\begin{align} 
 0=-(-\ddot{\rho}+2\ddot{\varphi}+\rho''-2\dot{\phi}^{2}-2\rho'\varphi'-2\dot{\rho}\dot{\varphi})f_{R}  \nonumber\\
+ \frac{M^{2}}{2}e^{2\rho}f+\frac{\partial^{2}}{\partial \tau^{2}}f_{R} \nonumber\\
 -\rho'\frac{\partial}{\partial x}f_{R}+\dot{\rho}\frac{\partial}{\partial \tau}f_{R} +\frac{q^{2}M^{2}e^{2\rho}}{2}    \nonumber\\
+e^{2\varphi}\left[-\frac{\partial}{\partial \tau}\left(e^{-2\varphi}\frac{\partial f_{R}}{\partial \tau} \right) +\frac{\partial}{\partial x} \left(e^{-2\varphi}\frac{\partial f_{R}}{\partial x} \right)    \right]
\end{align}

\begin{align} 
0=\frac{-M^{2}}{2}e^{2\rho}f-\left(\ddot{\rho}+2\varphi''-\rho''-2\varphi'^{2}-2\rho'\varphi'-2\dot{\rho}\dot{\varphi} \right)f_{R}  \nonumber\\
-\frac{q^{2}M^{2}e^{2\rho}}{2}+\frac{\partial^{2}}{\partial x^{2}}f_{R}-\dot{\rho}\frac{\partial f_{R}}{\partial \tau}
\nonumber  \\
-\rho' \frac{\partial f_{R}}{\partial x}-e^{2\varphi}\left[-\frac{\partial}{\partial \tau}\left( e^{-2\varphi}\frac{\partial f_{R}}{\partial \tau} \right) +\frac{\partial}{\partial x}\left(e^{-2\varphi}\frac{\partial f_{R}}{\partial x} \right) \right]  
\end{align}

\begin{equation} \label{eq3}
0=-(2\dot{\varphi}'-2\varphi'\dot{\varphi}-2\rho'\dot{\varphi}-2\dot{\rho}\varphi')f_{R}+\frac{\partial^{2}f_{R}}{\partial \tau \partial x}-\dot{\rho}\frac{\partial f_{R}}{\partial x}-\rho'\frac{\partial f_{R}}{\partial \tau}
\end{equation}

\begin{align} 
0=-2M^{2}e^{-2\varphi}f-e^{-2(\rho+\varphi)}(-\ddot{\varphi}+\varphi''+2\varphi'^{2}+2\dot{\varphi}^{2})f_{R}
\nonumber \\
+f_{R}  
+e^{-2(\rho+\varphi)}\left( \dot{\varphi}\frac{\partial f_{R}}{\partial t}-\varphi' \frac{\partial f_{R}}{\partial x}   \right)+\frac{q^{2}M^{2}e^{2\rho}}{2}
\nonumber  \\
-e^{-2\rho}\left[  -\frac{\partial}{\partial \tau}\left( e^{-2\varphi}\frac{\partial f_{R}}{\partial \tau}\right)+\frac{\partial}{\partial x} \left( e^{-2\varphi}\frac{\partial f_{R}}{\partial x}\right)  \right]
\end{align}

Now, let us consider perturbations around the background extremal solution as 
\begin{equation}
\label{rhovarphi}
\rho=-{\rm ln}({\rm cosh}\, x)+\delta \rho, \,\,\,\varphi=\delta \varphi
\end{equation}
The perturbed EoM are 
\begin{align} 
0=\frac{f_{R}(R_{0})+2M^{-2}f_{RR}(R_{0})}{2}\delta R  \nonumber\\
-f_{R}(R_{0})M^{-2}{\rm cosh}^{2}x (-\delta \ddot{\rho}+2\delta \ddot{\varphi}+\delta \rho''+2{\rm tanh}\, x\, \delta \phi')
\nonumber  \\
-2f_{R}(R_{0})M^{-2}\delta \rho+f_{RR}(R_{0})M^{-2}{\rm cosh}^{2}x({\rm tanh}\,x\, \delta R'+\delta R'')
\end{align}

\begin{align} 
0=-\frac{f_{R}(R_{0})+2M^{-2}f_{RR}(R_{0})}{2}\delta R +2f_{R}(R_{0})M^{-2}\delta \rho  \nonumber\\
-f_{R}(R_{0})M^{-2}{\rm cosh}^{2}\,x\,(\delta \ddot{\rho}+2\delta \varphi''-\delta \rho''+2{\rm tanh}\, x\,\delta \varphi')
 \nonumber \\
 +f_{RR}(R_{0})M^{-2}{\rm cosh}^{2}\,x\,({\rm tanh}\, x\,\delta R'+\delta \ddot{R})
\end{align}

\begin{align} 
0=-2(\delta \dot{\varphi}'+{\rm tanh}\, x\, \delta \dot{\varphi})\nonumber \\
+\frac{f_{RR}(R_{0})}{f_{R}(R_{0})}(\delta \dot{R}'+{\rm tanh}\, x\, \delta \dot{R})
\end{align}

\begin{align} 
0=-\frac{f_{R}(R_{0})+2M^{-2}f_{RR}(R_{0})}{2}\delta R \nonumber\\
-2M^{-2}f_{R}(R_{0})\delta \varphi \nonumber\\
-f_{R}(R_{0})M^{-2}\,{\rm cosh}^{2}\,x\, (-\delta \ddot{\varphi}+\delta \varphi'') \nonumber\\
-f_{RR}(R_{0})M^{-2}{\rm cosh}^{2}\,x\,(-\delta \ddot{R}+\delta R'')
\end{align}

A convenient parametrization of perturbations is 
\begin{equation}
\label{variations}
\delta \rho=\rho_{0}{\rm cosh}\, \omega \tau {\rm cosh}^{\beta}x,\,\,\,\,\delta \varphi=\varphi_{0}{\rm cosh}\, \omega \tau {\rm cosh}^{\beta}x
\end{equation}
where $\rho_{0},\phi_{0},\beta$ are arbitrary constants. 

Solving EoM, 
we find conditions 
\begin{equation}
\label{omegab}
\omega^{2}=\beta^{2}
\end{equation}
and 
\begin{equation}
\label{beta}
\beta=\beta_{\pm}=\frac{1}{2}\left[1\pm \sqrt{1-\frac{4}{3}M^{2}\left(\frac{f_{R}(R_{0})}{f_{RR}(R_{0})} \right)} \right]
\end{equation}
from 
\begin{equation}
\label{box}
\Box \delta \varphi=[\beta^{2}+\beta(\beta-1){\rm cosh}^{-2} \,x-\omega^{2}]\delta \varphi
\end{equation}
Let us note that 
$\beta$ has always a Real part which is positive, implying 
exponential instabilities.
In particular, for $\phi_{0}<0$ antievaporation
$\phi_{0}>0$ evaporation. 
Clearly, this is not enough to demonstrate that the extremal solution is unstable. 
So that, we show the numerical solution of the horizon radius 
obtained by 
EoM perturbed up to the second order in $\delta \rho,\delta \phi$.
Finally, we claim that a similar analysis in the case of the $SU(4)$-inspired 
model (despite of $SO(4)$ gauge group) leads to the same kind of instabilities,
as can be easily checked \footnote{
We mention that some solutions in other extended theories of gravity 
have also geodetic instabilities \cite{Addazi:2014mga}. 
}. 

\vspace{1cm}

\section{Conclusions}

In this paper, we have discussed dyonic BH solutions in $f(R)$-gravity coupled with 
a dilaton and two gauge bosons. 
We have shown that in the extremal limit of the internal radius saturating the 
external one, these solutions cannot be stable. 
In particular, their horizon radius explodes 
or rapidly decays in time, depending from the initial conditions. 
Such instabilities are interpreted as classical 
BH antievaporation and evaporation. 
Our result can be viewed as an analogous of the
Nojiri-Odintsov calculation in Ref. \cite{Nojiri:2014jqa}. 

The implications of our result 
are not fully understood. For instance, it is conceivable that 
(anti)evaporation can be related to deep theoretical issues
like the holographic conjecture \cite{Maldacena:1997re}, the no-remnant conjecture \cite{Susskind:1995da}
and, as a cascade, to many important related concepts 
of string theory and black holes. 
Such a result also represents an explicit violation of the 
generalized Birkhoff's theorem, as an example of unstable 
spherically symmetric solution. 

To conclude, 
in future companion papers, we hope to generalize our analysis in higher dimensions 
and in the case of extremal BH lying in a 
de Sitter or an Anti-de Siitter space-time.

\vspace{0.5cm} 

{\large \bf Acknowledgments} 
\vspace{3mm}

I also would like to thank the University of Tor Vergata (Roma) and The Republic (Coppito, L'Aquila) for the hospitality during the preparation of this paper. 
My work was supported in part by the MIUR research grant Theoretical Astroparticle Physics PRIN 2012CP-PYP7 and by SdC Progetto speciale Multiasse La Societa della Conoscenza in Abruzzo PO FSE Abruzzo 2007-2013.


\vspace{2cm}
\begin{thebibliography}{99}

\bibitem{Kallosh:1992ii}
  R.~Kallosh, A.~D.~Linde, T.~Ortin, A.~W.~Peet and A.~Van Proeyen,
  Phys.\ Rev.\ D {\bf 46} (1992) 5278
  doi:10.1103/PhysRevD.46.5278
  [hep-th/9205027].

\bibitem{Strominger:1996sh}
  A.~Strominger and C.~Vafa,
  Phys.\ Lett.\ B {\bf 379} (1996) 99
  doi:10.1016/0370-2693(96)00345-0
  [hep-th/9601029].
  
  \bibitem{Mathur:2009hf}
  S.~D.~Mathur,
  Class.\ Quant.\ Grav.\  {\bf 26} (2009) 224001
  doi:10.1088/0264-9381/26/22/224001
  [arXiv:0909.1038 [hep-th]].

  
\bibitem{Bianchi:2009ij}
  M.~Bianchi and M.~Samsonyan,
  Int.\ J.\ Mod.\ Phys.\ A {\bf 24} (2009) 5737
  [arXiv:0909.2173 [hep-th]].
  
\bibitem{Addazi:2015goa}
  A.~Addazi,
  Phys.\ Lett.\ B {\bf 757} (2016) 462
  doi:10.1016/j.physletb.2016.04.018
  [arXiv:1506.06351 [hep-ph]].
  
\bibitem{Addazi:2015yna}
  A.~Addazi, M.~Bianchi and G.~Ricciardi,
  JHEP {\bf 1602} (2016) 035
  doi:10.1007/JHEP02(2016)035
  [arXiv:1510.00243 [hep-ph]].

\bibitem{Addazi:2016mtn}
  A.~Addazi and M.~Khlopov,
  Mod.\ Phys.\ Lett.\ A {\bf 31} (2016) no.19,  1650111
  doi:10.1142/S021773231650111X
  [arXiv:1604.07622 [hep-ph]].

\bibitem{Addazi:2016xuh}
  A.~Addazi, J.~W.~F.~Valle and C.~A.~Vaquera-Araujo,
  Phys.\ Lett.\ B {\bf 759} (2016) 471
  doi:10.1016/j.physletb.2016.06.015
  [arXiv:1604.02117 [hep-ph]].



\bibitem{Ketov:2013dfa}
  S.~V.~Ketov and T.~Terada,
  JHEP {\bf 1312} (2013) 040
  doi:10.1007/JHEP12(2013)040
  [arXiv:1309.7494 [hep-th]].

\bibitem{Ferrara:2013pla}
  S.~Ferrara, A.~Kehagias and M.~Porrati,
  Phys.\ Lett.\ B {\bf 727} (2013) 314
  doi:10.1016/j.physletb.2013.10.027
  [arXiv:1310.0399 [hep-th]].



  
\bibitem{Hawking:1974sw}
  S.~W.~Hawking,
  Commun.\ Math.\ Phys.\  {\bf 43} (1975) 199
   Erratum: [Commun.\ Math.\ Phys.\  {\bf 46} (1976) 206].
  doi:10.1007/BF02345020


  

  
\bibitem{Bousso:1997wi}
  R.~Bousso and S.~W.~Hawking,
  Phys.\ Rev.\ D {\bf 57} (1998) 2436
  doi:10.1103/PhysRevD.57.2436
  [hep-th/9709224].
  
\bibitem{Nojiri:2013su}
  S.~Nojiri and S.~D.~Odintsov,
  Class.\ Quant.\ Grav.\  {\bf 30} (2013) 125003
  doi:10.1088/0264-9381/30/12/125003
  [arXiv:1301.2775 [hep-th]].
  
\bibitem{Nojiri:2014jqa}
  S.~Nojiri and S.~D.~Odintsov,
  Phys.\ Lett.\ B {\bf 735} (2014) 376
  doi:10.1016/j.physletb.2014.06.070
  [arXiv:1405.2439 [gr-qc]].
  
   
\bibitem{Addazi:2016prb}
  A.~Addazi and S.~Capozziello,
  Mod.\ Phys.\ Lett.\ A {\bf 31} (2016) no.09,  1650054
  doi:10.1142/S0217732316500541
  [arXiv:1602.00485 [gr-qc]].
  
\bibitem{Addazi:2014mga}
  A.~Addazi and S.~Capozziello,
  Int.\ J.\ Theor.\ Phys.\  {\bf 54} (2015) no.6,  1818
  doi:10.1007/s10773-014-2387-z
  [arXiv:1407.4840 [gr-qc]].
  
  
  \bibitem{Maldacena:1997re}
  J.~M.~Maldacena,
  Int.\ J.\ Theor.\ Phys.\  {\bf 38} (1999) 1113
   [Adv.\ Theor.\ Math.\ Phys.\  {\bf 2} (1998) 231]
  doi:10.1023/A:1026654312961
  [hep-th/9711200].
  
    
  \bibitem{Susskind:1995da}
  L.~Susskind,
  hep-th/9501106.



 
\end{thebibliography}
\end{document}